\documentclass[review]{elsarticle}
\usepackage{amsmath}
\usepackage{amsthm}
\usepackage{amssymb}
\usepackage{titlesec}
\usepackage[thinlines,thiklines]{easybmat}

\usepackage{graphicx}
\theoremstyle{definition}\newtheorem{Df}{Definition}
\theoremstyle{plain}\newtheorem{Th}{Theorem}
\theoremstyle{definition}\newtheorem{Rm}{Remark}
\theoremstyle{definition}
\theoremstyle{plain}
\theoremstyle{plain}\newtheorem{Co}{Corollary}
\theoremstyle{plain}\newtheorem{Lm}{Lemma}
\theoremstyle{plain}

\begin{document}
\begin{frontmatter}
\title{  On  hybrid models of quantum finite automata }
\author[a1,a2]{Lvzhou Li\corref{one}}
\author[a2]{Yuan Feng}

 \cortext[one]{Corresponding author.\\
 \indent {\it E-mail
address:}  lilvzh@mail.sysu.edu.cn (L.
Li); Yuan.Feng@uts.edu.au (Y. Feng)}

\address[a1]{Department of
Computer Science, Sun Yat-sen University, Guangzhou 510006,
  China}

\address[a2]{Center for Quantum Computation and Intelligent
Systems, FEIT, University of Technology Sydney, Sydney 2oo7, Australia}

\begin{abstract}

 In the literature, there exist several quantum finite automata (QFA) models with both quantum and classical states.  These  models are of particular interest,  as they show  praiseworthy advantages over the fully quantum models in some nontrivial aspects.
This paper characterizes these models in a uniform framework by proposing a general hybrid model consisting of a quantum component and a classical one which can interact with each other. The existing hybrid QFA can be naturally regarded as the general model with specific communication patterns (classical-quantum, quantum-classical, and two-way, respectively). We further clarify the relationship between these hybrid QFA and some other quantum models. In particular, it is shown that hybrid QFA can be simulated exactly by QFA with quantum operations, which in turn has a close relationship with two early proposed models: {\it ancialla QFA}  and {\it quantum sequential machines}.

\end{abstract}

\end{frontmatter}

\section{Introduction}

Quantum finite automata (QFA), as a theoretical model for quantum
computers with finite memory,  have interested many researchers (see  e.g. [1-7, 9-17, 19,20, 22-25,29]).
  So far,  a variety of models of QFA have been introduced and explored to
various degrees (one can refer to a review aticle  \cite{QLMG12} and references therein).  Roughly speaking, these QFA models fall into the following two categories: one-way QFA (1QFA), where the tape head is required to move  right on scanning each tape cell, and two-way QFA (2QFA), where the tape head is allowed to move left or right, and even to stay stationary. Notably, 2QFA are strictly more powerful than 1QFA: the former is able to recognize \footnote{In this paper, recognizing a language always means recognizing it with bounded error.} non-regular languages \cite{KW97}, while the latter only regular ones \cite{BC01,BP99,KW97}.

Another criterion which is used to classify different QFA is  the state evolution type.
In  early references, the state evolution of a QFA is  assumed to be  unitary operators, in accord with the postulate of quantum mechanics that the state evolution of a closed quantum system is described by a unitary
transformation. Later on, it was realized that a QFA does not need to be a closed system; it  can interact with the environment. Thus the  state evolution  should be   general quantum operations, i.e., trace-preserving completely positive mappings (see  Hirvensalo \cite{Hir08a, Hir08b}, Li et al \cite{LQ12} and Yakaryilmaz et al \cite{YC10}).
QFA with quantum operations have been thought to be a nice definition for QFA, since they possess nice closure properties and have a competitive computational power with their classical counterparts.

Another model worth mentioning  is the  {\it ancilla QFA}  proposed in \cite{Pas00}. Actually, ancilla QFA  represent the same model as 1QFA with quantum operations, but  in different forms, which will become  clear in later sections.
Interestingly,  ancilla QFA can  also be regarded as  {\it quantum sequential machines} \cite {Qiu02, LQ06}, assigned with some accepting states.  The relationship between these models will be elaborated in this paper.

\subsection{Hybrid models of QFA}
 In the literature, there is a class of QFA that differ from other  QFA models  by consisting of two interactive components: a quantum component and a classical one. We call them {\it hybrid models of QFA} in this paper. These hybrid models are of particular interest,  as they show  praiseworthy advantages over the fully quantum models in some nontrivial aspects.

The first hybrid model of QFA is the {\it two-way quantum finite automata with quantum and classical states} (2QCFA, for short) proposed by Ambainis and Watrous \cite{AW02} in which on scanning an input symbol, the internal states evolve as follows: first the quantum part undergoes a unitary operator or a projective measurement that is  determined by the current classical state and the scanned symbol, and then the classical part specifies a new classical state and a movement of the tape head (left, right, or stationary), which depends on the scanned symbol (along with the outcome if the quantum part makes a measurement).  In  \cite{AW02}, it was shown that 2QCFA  are strictly more powerful than their classical counterparts---two-way probabilistic finite automata (2PFA). In this paper we will discuss a one-way variant of 2QCFA, denoted by 1QCFA (where the tape head is restricted to move towards only right).

Another hybrid model in the literature is the  {\it one-way quantum finite automata with control language} (CL-1QFA, for short) proposed by Bertoni \cite{Ber03} in which on scanning an input symbol, a unitary operator followed by a projective measurement is applied to the current quantum state. Thus, a CL-1QFA fed with an input string $x$ finally produces a sequence $y$ of measurement results with a certain probability. The input string $x$ is said to be accepted if $y$ is in a given regular language (i.e., the control language). It was shown in \cite{Ber03, MP06} that CL-1QFA recognize exactly the class of regular languages, and CL-1QFA can be more succinct (i.e.,  have less states) than deterministic finite automata (DFA)  for certain languages.

Very recently, Qiu et al \cite{QMS09} proposed a new hybrid model called {\it 1QFA together with classical states} (1QFAC, for short) which consists of a quantum component and a classical one.  On scanning an input symbol, the two components interact  in the following way: first the quantum component undergoes a unitary operator that is determined by the current classical state and the scanned symbol, and then the classical component evolves like a DFA.
After  scanning the whole input string, a projective measurement determined by the final classical state is performed on the final quantum state, giving the accepting and rejecting probabilities.
In \cite{QMS09},  it was proved that  1QFAC recognize all regular languages and  are exponentially more concise than DFA for certain languages.

\subsection{Motivation and contribution of the paper}
The hybrid models mentioned above  are of particular interest and worthy of further consideration, at least for the following  reasons. First, hybrid models are easier to be physically implemented than fully quantum models. For example, while the position of the tape head in the 2QFA model introduced by Kondacs and Watrous \cite{KW97} is a superposed quantum state and at least $O(\log n)$ qubits are required to store it (where $n$ is the length of the input), in a 2QCFA the tape head position is merely a classical variable which is easy to store and manipulate.  Second, hybrid models often save states by being augmented with a small number of quantum states. For example,  CL-1QFA and 1QFAC have been proved to be much smaller than DFA when accepting the same language. Therefore, it is beneficial to construct a hybrid model for a practical problem (language) with an appropriate  trade-off between quantum and classical states.
 Last but not least,  quantum engineering systems developed in the further
will most probably have a classical human-interactive interface and a quantum processor, and thus they are  hybrid models.  In fact, hybrid models have already been  encountered several times in quantum computing, varying from quantum Turing machines  \cite{Wat03} and quantum
finite automata \cite{AW02,Ber03,QMS09} to multiple-prover quantum interactive proof systems  (the model of complexity class MIP$^*$ \cite{CHTW}) and  quantum programming \cite{Sel04}.\footnote{As
stated by Selinger \cite{Sel04}, quantum programs can be described by ``quantum data
with classical control flows''. Thus, a quantum program can be viewed as a hybrid model where quantum data are represented by states of a
quantum component and classical control flows are implemented by a classical
component.}

\begin{figure}[tbp]\centering
\includegraphics[width=65mm]{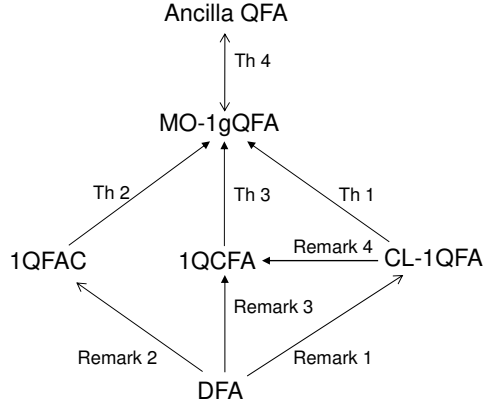}
\caption{ A diagram illustrating the relationship between the models mentioned in this paper. A one-directional arrow denotes that the model at the tail can be simulated by the one at the head; and a bidirectional arrow means that the two models  can simulate each other.  Note that all  models in the diagram have the same language recognition power,
recognizing   the class of regular languages.}\label{Fig-relation}
\end{figure}

 This paper aims  to characterize the structure of these hybrid models in a uniform framework and to  clarify the relationship between these models and others.
The contribution of this paper is as follows. (i) First, we  characterize three  hybrid models of QFA (1QCFA, CL-1QFA, and 1QFAC ) in a uniform framework: each hybrid QFA is represented as a communication system consisting of a quantum component and a classical one with the communication between  them modeled by  controlled operations.  The three models differ from each other only in the specific communication pattern.
(ii) Second,  we will show that  1QCFA,  CL-1QFA, and 1QFAC can all be simulated exactly by 1QFA with quantum operations.
1QFA with quantum operations actually represent the same model as another early proposed model---ancilla QFA, which in turn can be  regarded as quantum sequence machines assigned with some accepting states. We refer to Fig. \ref{Fig-relation} for their detailed relationship.

\subsection{Organization of the paper}
The reminder of this paper is organized as follows. Some preliminaries from quantum theory, automata theory and controlled operations are presented in Section \ref{sec:pre}.  In Section \ref{sec:hm} we characterize the structure of three hybrid models of QFA in a uniform framework. Section \ref{sec:relation}  clarifies the relationship between hybrid models of QFA and other models. Some results concerning the language recognition power and the equivalence problem of hybrid QFA follow as corollaries there.  Some conclusions are made in Section  \ref{sec:con}.

\section{Preliminaries}\label{sec:pre}

\subsection{Preliminaries from  quantum theory}\label{sec:pre-qt}
For convenience of the reader we briefly recall some basic notions from quantum theory. We refer to \cite{NC00} for more details.
According to von Neumann's formalism of quantum mechanics, a quantum system is associated with a Hilbert space which is called the state space of the system. In this paper, we only consider finite dimensional spaces. A (mixed) state of a quantum system is represented by a density operator on its state space. Here a density operator $\rho$ on ${\cal H}$ is a positive semi-definite linear operator such that $\text{Tr}(\rho)= 1$. When the rank of $\rho$ is $1$, that is, $\rho = |\psi\rangle\langle\psi|$ for some $|\psi\rangle\in {\cal H}$, then $\rho$ is called a pure state. Let $L({\cal H})$ and $D({\cal H})$ be the sets of linear operators and density operators on ${\cal H}$, respectively.

The evolution of a closed quantum system is described by
a unitary operator $U$ on its state space. If the states of the system at times $t_1$
and $t_2$ are
$\rho_{1}$ and $\rho_{2}$, respectively, then $\rho_{2} = U \rho_{1} U^{\dag}$ for some unitary operator $U$ which depends only on $t_{1}$ and $t_{2}$. Here $U^\dag$ is the complex conjugate and transpose of $U$.

In contrast, the evolution of an \emph{open} quantum system is
characterized by  a quantum
operation ${\cal E}$ on its state space ${\cal H}$, which is a linear map from $L(\mathcal{H})$ to itself that has an {\it operator-sum
representation} as
\begin{align}
{\cal E}(\rho)=\sum_kE_k\rho E_k^\dagger,\label{OP}
\end{align}
where
$\{E_k\}$, known as the {\it operation elements} of ${\cal E}$, are linear operators
on  ${\cal H}$. Furthermore, ${\cal E}$
is said to be {\it trace-preserving} if the following completeness condition is satisfied:
\begin{align}
\sum_kE_k^\dagger E_k=I.\label{OPC}
\end{align}
 Throughout the rest of this paper, when referring to a quantum operation ${\cal E}$, it is always assumed to be trace-preserving.

To extract  information from a quantum system, a   measurement has to be performed.  A general measurement is described by a collection $\{M_m\}$ of measurement operators,
where the index $m$ refers to the potential measurement outcome,  satisfying the completeness condition
 \begin{align*}
\sum_mM_m^\dagger M_m=I.
\end{align*}
If this measurement is performed on  a state $\rho$, then the classical outcome $m$ is obtained with the probability
$p(m)=\text{Tr}(M_m^\dagger M_m\rho)$, and the post-measurement state is
\begin{equation*}\frac{M_m\rho M_m^\dagger}{\sqrt{p(m)}}.\end{equation*}
For the case that $\rho$ is a pure state $|\psi\rangle$, that is, $\rho = |\psi\rangle\langle \psi|$, we have
$p(m)=\|M_m|\psi\rangle\|^2,$
and the state $|\psi\rangle$   ``collapse'' into the state
 \begin{equation*}\frac{M_m|\psi\rangle}{\sqrt{p(m)}}.\end{equation*}

A special case of general measurements is the projective measurement $\{P_m\}$ where $P_m$'s are orthogonal projectors.

Suppose we have physical systems $A$ and $B$, whose state is described by a density operator $\rho^{AB}\in D({\cal H}_A \otimes {\cal H}_B)$. Then the state for system $A$ is $\rho^A=\text{Tr}_B(\rho^{AB})$ where
$\text{Tr}_B$, known as  the {\it partial trace } over system $B$,  is defined by  a linear map from ${\cal H}_A\otimes {\cal H}_B$ to ${\cal H}_A$ satisfying$$\text{Tr}_B(|a_1\rangle \langle a_2|\otimes |b_1\rangle\langle b_2|)=\langle b_2|b_1\rangle |a_1\rangle \langle a_2|$$ where $|a_1\rangle, |a_2\rangle\in {\cal H}_A$, and $|b_1\rangle, |b_2\rangle\in {\cal H}_B$.

\subsection{Preliminaries from automata theory}

 For a non-empty set $\Sigma$, denote by $\Sigma^{k}$ and $\Sigma^{*}$ the
sets of all strings over $\Sigma$ with length $k$ and with finite length, respectively. Let $|w|$ be the
length of string $w$.

A DFA is a five-tuple ${\cal A}=(S,\Sigma,s_1,\delta,S_a)$ where $S$ is a finite state set, $\Sigma$ is a finite alphabet, $s_1\in S$ is the initial state, $S_a\subseteq S$ is the accepting set, and $\delta: S\times\Sigma\rightarrow S $ is the transition function\footnote{Without loss of generality, $\delta$ is required to be a total function.}: $\delta(s,\sigma)=t$ means that the current state $s$  changes to $t$ when scanning $\sigma$. Furthermore, $\delta$ can be extended to $\delta^*: S\times\Sigma^*\rightarrow S$ by defining:
 i) $\delta^*(s,\epsilon)=s$, and ii) $\delta^*(s,x\sigma)=\delta(\delta^*(s,x),\sigma)$ where $x\in\Sigma^*$ and $\sigma\in \Sigma$. ${\cal A}$ is said to accept $x\in\Sigma^*$, if $\delta^*(s_1,x)\in S_a$.

We can also describe a DFA using the  matrix notation. Let $A_\sigma$ be an $|S|\times |S|$ matrix with $A_{\sigma}[i,j]$ being $1$ if $\delta(s_j, \sigma) = s_i$, and $0$ otherwise. Then each column of $A_{\sigma}$  has a unique $1$ and other entries are all 0. Let $\pi$ be a 0-1 column vector with only the first entry being $1$, and $\eta$ the 0-1 row vector with $\eta_i=1$ iff $s_i\in S_a$. Define a function $f_A: \Sigma^*\rightarrow \{0,1\}$ as
\[f_A(x)=\eta A_{x_{|x|}}\cdots A_{x_2}A_{x_1}\pi.\]
Then $f_A(x)=1$ iff $A$ accepts $x$.

\subsection{Controlled operations}\label{sec:co}

Let ${\cal H}_{A}\otimes {\cal H}_{B}$  be the state space of a bipartite quantum system $AB$. Let $\{P_i\}_{i=1}^n$ be a projective measurement on ${\cal H}_{A}$ with $n$ outcomes, and $\mathcal{E}_i$, $1\leq i\leq n$, quantum operations on ${\cal H}_{B}$. The \emph{controlled operation} {\bf`if subsystem $A$ was measured in result $i$, then $\mathcal{E}_i$  is performed on subsystem $B$'} can be defined by the following operation elements
\begin{align}
\{P_i\otimes E_k^i:i=1,\cdots, n, k\in K_i\} \label{CO}
\end{align}
where $\{E^i_k\}_{k\in K_i}$ are operation elements of $\mathcal{E}_i$, that is, $\mathcal{E}_i(\rho)=\sum_{k\in K_i} E^i_k\rho {E^i_k}^\dagger$. It is straightforward to check  that the elements in (\ref{CO}) satisfy the completeness condition.
Furthermore, for $\rho\otimes\varrho \in L({\cal H}_{A}\otimes {\cal H}_{B} )$, we have
\[\mathcal{E}(\rho\otimes\varrho)=\sum_{i=1}^n P_i\rho P_i\otimes\mathcal{E}_i(\varrho).\]

If the  projective measurement $\{P_i\}_{i=1}^n$ in the above is replaced by a general measurement $\{M_i\}_{i=1}^n$, then we get a  quantum operation $\mathcal{E}'$ that has the following effect
\[\mathcal{E}'(\rho\otimes\varrho)=\sum_{i=1}^n M_i\rho M_i^\dagger\otimes\mathcal{E}_i(\varrho).\]

\section{Characterization of structures of hybrid models of QFA}\label{sec:hm}
In this section,   structures of several existing hybrid models of QFA are characterized in a uniform framework: a hybrid QFA can be regarded as a communication system consisting of two blocks, with the communication between them  modeled by controlled operations, which thus provides an intuitive insight into the structure of hybrid models.

\subsection{ Hybrid models: the general definition}\label{subsec:6}

\begin{figure}[htbp]\centering
\includegraphics[width=60mm]{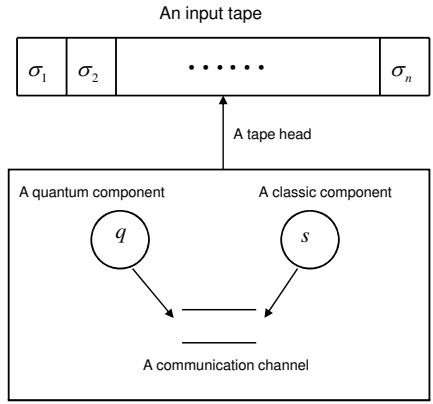}
\caption{ A diagram illustrating the idea behind  hybrid models of QFA.  }\label{Model}
\end{figure}

Hybrid models discussed in this paper are depicted  in Fig. \ref{Model}. A hybrid model of QFA comprises a quantum component, a classical component, a classical communication channel, and a classical tape head (that is, the tape head is regulated by the classical component). On scanning an input symbol, the quantum and classical components interact to evolve into new states, during which communication may occur between them.  In this paper, we focus on hybrid models with a one-way tape head, that is, on scanning an input symbol the model moves its tape head one cell right.

In the literature there are three  models of QFA (CL-1QFA, 1QFAC, and 2QCFA) fitting into  Fig. \ref{Model}.
For the last model, we  will consider its on-way variant  1QCFA. We will show that these models  can be characterized in a uniform framework as in  Fig. \ref{Model}, with the  only difference being the specific communication pattern:
\begin{itemize}
  \item In CL-1QFA, only quantum-classical communication is allowed, that is, the quantum component sends its measurement result to the classical component, but no reverse communication is permitted.
  \item In 1QFAC, only classical-quantum communication is allowed, that is, the classical component sends its current state to the quantum component.
  \item In 1QCFA,  two-way communication is allowed: (1) first,  the classical component sends its current state to the quantum component; (2) second, the quantum component sends its measurement result to the classical component.
\end{itemize}

Throughout the remainder of this paper we let ${\cal H}_Q$, where $Q=\{q_1,q_2,\cdots,q_n\}$,  be the n-dimensional  Hilbert space  spanned by the orthonormal vectors $\{|q_i\rangle: i=1,\cdots,n\}$. Mathematically, $|q_i\rangle$ is an $n$-dimensional column vector having $1$ as the $i$th entry and $0$ else.  Sometimes we abuse the notation slightly by writting $q_i$ directly for $|q_i\rangle$.

\subsection{CL-1QFA: a hybrid model with quantum-classical communication }\label{sec:cl1qfa}

Bertoni {\it et al} \cite{Ber03} introduced a   QFA model  called {\it one-way QFA with control language} (CL-1QFA),  defined as follows.
\begin{Df}
A  CL-1QFA  is a 7-tuple $${\cal
  A}=(Q, \Sigma, {\cal C},  q_1, \{U_\sigma\}_{\sigma\in\Sigma}, \mathcal{M}, {\cal L}),$$ where
 $Q$ is a finite state set, $\Sigma$ is a finite alphabet,
${\cal C}$ is a finite set of symbols  (measurement outcomes), $q_1\in Q$ is the initial quantum state, $U_{\sigma}$ is a unitary operator for each $\sigma\in\Sigma$,    $\mathcal{M}$ is a projective measurement given by a collection $\{P_c\}_{c\in{\cal C}}$ of projectors,
and ${\cal L}\subseteq{\cal C}^{*}$ is a regular language
(called a control language).\label{Df:CL-QFA}
\end{Df}

In CL-1QFA ${\cal A}$, on scanning  a symbol $\sigma$, a unitary operator $U_{\sigma}$ followed by the projective measurement $\mathcal{M}$ is performed on its current state. Thus, given an input string $x\in\Sigma^*$, the computation produces a
sequence $y\in {\cal C}^{*}$  of measurement results with a
certain probability $p(y|x)$ that is  given by
\begin{align}
p(y_1\dots y_{n}|x_1\dots
x_n)=\left\|\prod^{n}_{i=1}(P_{y_i}U_{x_i})|q_1\rangle\right\|^2
\end{align}
 where we define  the ordered product $\prod_{i=1}^{n}A_i= A_nA_{n-1}\cdots A_1$. The computation is said to be {\it accepted } if $y$ belongs to a fixed regular  language ${\cal
L}\subseteq {\cal C}^{*}$.
Thus the probability of ${\cal M}$
accepting $x$ is
\begin{align}
P_{\cal A}(x)=\sum_{y_1\dots y_{n}\in {\cal L}}p(y_1\dots
y_{n}|x_1\dots x_n).\label{f_CL}
\end{align}

\begin{figure}[tbp]\centering
\includegraphics[width=113mm]{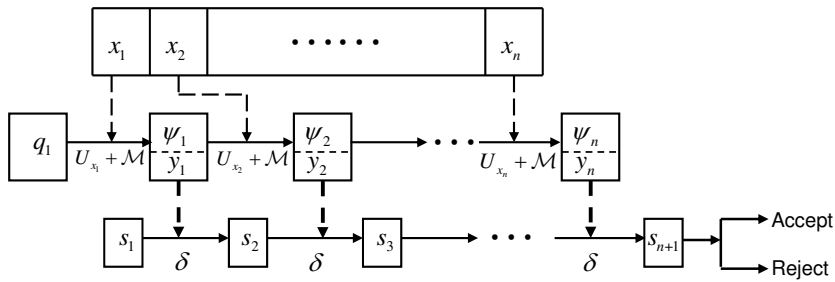}
\caption{A CL-1QFA is  a hybrid model with one-way communication allowed from the quantum component to the classical one,  indicated by thick dashed lines. Here $\delta$ is the transition function of the DFA accepting ${\cal L}$.  }\label{Fig-CL1QFA}
\end{figure}

 Obviously, a CL-1QFA  can be regarded as a hybrid model comprising  the following two components:
\begin{itemize}
  \item A  quantum component with state space ${\cal H}_Q$ that undergos unitary operators  $\{U_\sigma\}$ and projective measurement ${\cal M}$.
  \item  A classical component that is the DFA accepting the control language ${\cal L}$.
\end{itemize}
Note that  the classical DFA accepting  ${\cal L}$ takes the measurement outcomes of the quantum component as input.
Thus, communication occurs from the quantum component to the classical one. The structure of  CL-1QFA is illustrated in Fig. \ref{Fig-CL1QFA}.

\begin{Rm}
It was proved in \cite{MP06} that for each regular language, there exists a CL-1QFA recognizing it with certainty. The idea can be roughly described in  Fig.  \ref{Fig-CL1QFA} as follows.  Given a regular language $L$, first the quantum component  is elaborately designed to function as a bijective mapping from $\Sigma^*$ to $\mathcal{C}^*$, such that $\mathcal{L}\subseteq\Sigma^*$ is mapped to a regular language $L'\subseteq \mathcal{C}^*$. Then the classical component is designed to be a DFA accepting $L'$. One can refer to \cite{MP06} for more details.\label{rm:cl1qfa}
\end{Rm}

\subsection{1QFAC: a hybrid model with classical-quantum communication } \label{sec:1qfac}
Recently, Qiu et al \cite{QMS09} proposed a new  model named {\it 1QFA together with classical states} (1QFAC), defined as follows.\footnote{In this paper we  consider only the case that 1QFAC are language acceptors, and one can refer to \cite{QMS09} for a more general definition.}

\begin{Df}
A 1QFAC ${\cal A}$ is defined by a 8-tuple
$${\cal A}=(Q, S, \Sigma, q_1, s_{1}, \{ U_{s,\sigma}\}_{s\in S, \sigma\in \Sigma} , \delta,
 \{ {\cal M}_s\}_{s\in S}),$$
where  $Q$ and $S$ are finite sets of quantum states and classical states, respectively, $\Sigma$  is a finite input alphabet,
$q_1\in Q$ and $s_1\in S$ are initial quantum and classical states, respectively,  $U_{s,\sigma}$ is a unitary operator on ${\cal H}_Q$ for each $s$ and $\sigma$,
  $\delta: S\times \Sigma\rightarrow S$ is a  classical transition  function, and for each $s$, ${
\cal M}_s$ is a projective measurement given by projectors $\{P_{s,a}, P_{s,r}\}$ where the two outcomes $a$ and $r$ denote accepting and rejecting, respectively.
\end{Df}

The machine starts with the initial states $s_{1}$ and $q_1$.
 On scanning an input symbol $\sigma\in\Sigma$,  $U_{s,\sigma}$ is first applied to the current quantum state, where $s$ is the current classical state; afterwards, the  classical state $s$ changes to $t=\delta(s,\sigma)$.
Finally, when the whole input string is finished,    a measurement  ${\cal M}_s$ determined by the final classical state is performed on the final quantum state, and the input is accepted if the outcome $a$ is observed.  Therefore, the probability of 1QFAC ${\cal A}$ accepting $x=x_1x_2\cdots x_n\in\Sigma^*$ is given by
\begin{align}P_{\cal A}(x)=\|P_{s_{n+1}, a}U_{s_n,x_n}\cdots U_{s_2,x_2}U_{s_1,x_1}|q_1\rangle\|^2 \label{prob:1qfac}\end{align}
where $s_{i+1}=\delta(s_i,x_i)$ for $i=1,\cdots, n$.

\begin{figure}[tbp]\centering
\includegraphics[width=115mm]{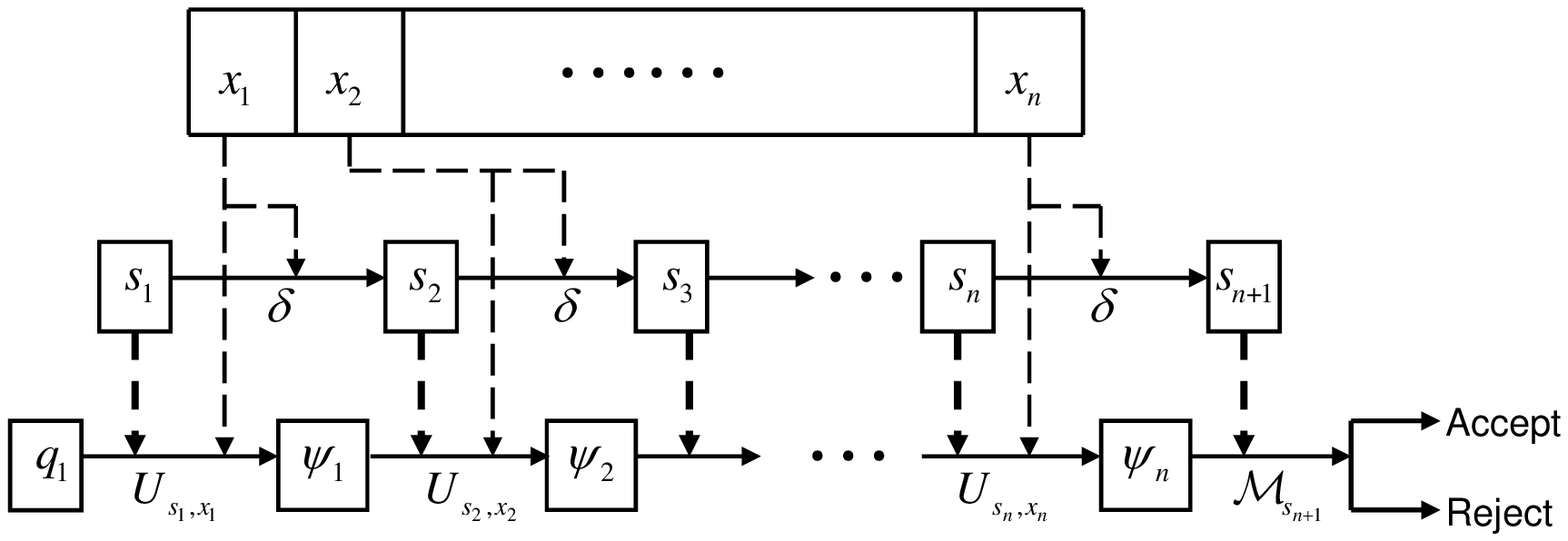}
\caption{A 1QFAC is a hybrid model with one-way communication allowed from the classical component to the quantum one, indicated by thick dashed lines.    }\label{Fig-1QFAC}
\end{figure}

Again, it is easy to see that a 1QFAC is actually a hybrid model comprising the follows two components:
\begin{itemize}
  \item A  quantum component with state space ${\cal H}_Q$, undergoing unitary operators $\{U_{s,\sigma}\}$.
  \item A classical component represented by a DFA ${\cal A}'=(S, \Sigma, s_1, \delta)$ without an accepting set.
\end{itemize}
Note that each unitary operator $U_{s,\sigma}$ is determined by the current classical state $s$. Thus, communication is required from the classical component to the quantum one. The structure of 1QFAC is illustrated in Fig. \ref{Fig-1QFAC}.

\begin{Rm}It follows straightforward from Fig. \ref{Fig-1QFAC} that for each regular language $\mathcal{L}$, there exists a 1QFAC recognizing it with certainty. For that, the classical component is designed to be a DFA accepting $\mathcal{L}$,  and the quantum component is set to be a qubit system with the orthnormal basis $|0\rangle$ and $|1\rangle$, with $|1\rangle$ being the initial state. Each operator $U_{s,\sigma}$ is simply set to be $I$. If $s$ is an accepting state, then let $P_{s,a}=|1\rangle\langle 1|$ and $P_{s,r}=|0\rangle\langle 0|$; otherwise, let $P_{s,a}=|0\rangle\langle 0|$ and $P_{s,r}=|1\rangle\langle 1|$.
\end{Rm}

\subsection {1QCFA: a hybrid model with two-way communication}\label{sec:1QCFA}

Ambainis and Watrous \cite{AW02} proposed the model of {\it two-way QFA with quantum and classical states} (2QCFA).  As proved in \cite{AW02}, 2QCFA  can recognize
non-regular language $L_{eq}=\{a^{n}b^{n}|n>0\}$ in polynomial time and  the palindrome language $L_{pal}=\{x\in
\{a,b\}^{*}|x=x^{R}\}$ in exponential time, which shows the superiority of 2QCFA over their classical counterparts 2PFA.   In the following we  discuss 1QCFA, the one-way variant of 2QCFA.
\begin{Df}
A 1QCFA  is specified by a 9-tuple
$$
\mathcal{A}=(Q,S,\Sigma, {\cal C}, q_1, s_{1}, \{\Theta_{s,\sigma}\}_{s\in S,\sigma\in\Sigma},\delta,S_a)
,$$
where
$Q$ and  $S$  are finite sets of quantum and classical states, respectively, $\Sigma$ is a finite input alphabet,  ${\cal C}$ is a finite set of symbols (measurement outcomes), $q_1\in Q$ and  $s_1\in S$  are initial quantum and classical states, respectively,     $\Theta_{s,\sigma}$  for each $s$ and $\sigma$ is a  general measurement  on ${\cal H}_Q$ with outcome set ${\cal C}$, $\delta: S\times \Sigma\times C\rightarrow S$ specifies the classical state transition, and $S_a\subseteq S$ denotes a set of accepting states.\label{Df-1QCFA}
\end{Df}

The notion of 1QCFA given above is  slightly more general than the one-way version of  2QCFA in \cite{AW02},
where each $\Theta_{s,\sigma}$ is required to be either a unitary operator or a projective measurement, both being special cases of general measurement considered here.

On scanning a symbol $\sigma$, first  the general measurement $\Theta_{s,\sigma}$ determined by the current classical state $s$ and the scanned symbol $\sigma$ is performed on the current quantum state,  producing some outcome $c\in {\cal C}$; then the classical
 state  $s$ changes to $s'=\delta(s, \sigma, c)$ by reading $\sigma$ and $c$. After scanning all input symbols, ${\cal A}$ checks whether its classical state is in $S_a$. If yes,  the input is accepted; otherwise, rejected.  Therefore, the probability of 1QCFA ${\cal A}$ accepting $x=x_1x_2\cdots x_n\in\Sigma^*$ is given by
\begin{align}P_{\cal A}(x)=\sum_{c_1c_2\cdots c_n\in{\cal C}^n}\chi_a(s_{n+1})\|M^{c_n}_{s_n,x_n}\cdots M^{c_2}_{s_2,x_2}M^{c_1}_{s_1,x_1}|q_1\rangle\|^2 \label{prob:1qfac}\end{align}
where:
\begin{itemize}
  \item [(1)] $\chi_a: S\rightarrow\{0,1\}$ is defined by  $\chi_a(s)=\left\{
                                                                \begin{array}{ll}
                                                                  1, & \hbox{if $s\in S_a$;} \\
                                                                  0, & \hbox{otherwise.}
                                                                \end{array}
                                                              \right.
  $
  \item  [(2)] $\{M^{c}_{s,\sigma}\}_{c\in {\cal C}}$ are measurement operators of $\Theta_{s, \sigma}$.
  \item  [(3)] $s_{i+1}=\delta(s_i,  x_i, c_i)$ for $i=1,\cdots, n$.
\end{itemize}

\begin{figure}[htbp]\centering
\includegraphics[width=117mm]{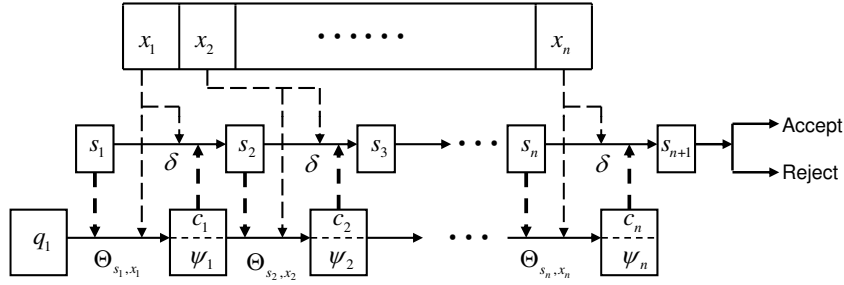}
\caption{A 1QCFA is  a hybrid model consisting of  a quantum component and a classical one with two-way communication, indicated by thick dashed lines. }\label{Fig-1QCFA}
\end{figure}

Obviously, a 1QCFA is a hybrid model comprising the following two components:
\begin{itemize}
  \item A quantum component with state space ${\cal H}_Q$, undergoing general measurements $\{\Theta_{s,\sigma}\}$ with outcome set ${\cal C}$.
  \item A classical component represented by a DFA ${\cal A}'=(S,\Sigma\times {\cal C}, s_1, \delta, S_a)$.
\end{itemize}
Note that two-way communication occurs between the two components, since each $\Theta_{s,\sigma}$ is determined by the classical state $s$ and the DFA ${\cal A}'$ has to scan the outcome $c$ produced by the quantum component.
The structure of  1QCFA is illustrated in Fig. \ref{Fig-1QCFA}.

\begin{Rm}
From Fig. \ref{Fig-1QCFA} it is easy to see that for each regular language $\mathcal{L}$, there exists a 1QCFA recognizing it with certainty.  Actually, the classical component alone is already sufficient for the task.
\end{Rm}

\begin{Rm}From Fig. \ref{Fig-1QCFA}  one can also see that a 1QCFA reduces to a CL-1QFA, if the following restrictions are made: \begin{itemize}
      \item [(i)] Each  $\Theta_{s,\sigma}$ has the form of a unitary operator $U_\sigma$  followed by a projective measurement ${\cal M}$. Thus no communication is required from the classical component to  the quantum one.
        \item [(ii)] The classical transition $\delta$ is independent of  $\sigma$.
                                                                                                                               \end{itemize}
 As a result, CL-1QFA are included as a subset in 1QCFA. However, as will be shown in the next section, these two models actually accept the same set of languages. Thus, the communication from the classical component to the quantum one seems to be superfluous. We believe that two-way communication is useful for reducing the number of states, which, however, needs to be  further explored.

In addition,  1QFAC illustrated in Fig. \ref{Fig-1QFAC} can also be regarded as a variant of 1QCFA in Fig. \ref{Fig-1QCFA} by restricting all $\Theta_{s,\sigma}$ to be unitary operator $U_{s,\sigma}$ (thus no communication is needed from the quantum part to the classical one) and  changing the acceptance fashion.
\label{Rm:4}
\end{Rm}

\section{Relationship between QFA models}\label{sec:relation}
Based on the characterization of structures of CL-1QFA, 1QFAC, and 1QCFA in the previous section,  we will prove in this section that these models can all be simulated exactly \footnote{A QFA simulating another one exactly means that they have the same accepting probability for each input string.} by the model of 1QFA with quantum operations studied in \cite{Hir08a, Hir08b, LQ12, YC10}. For simplicity, throughout the rest of this paper we will use the abbreviation  MO-1gQFA (measured-once one-way general QFA) to denote the model of 1QFA with quantum operations, in accordance with the name used in \cite{LQ12}. After clarifying the relationship between hybrid models and MO-1gQFA, we will also reveal the relationship between MO-1gQFA and two early proposed models---ancialla QFA and quantum sequential machines.

\subsection{Review of MO-1gQFA}
In the following, we  recall the definition of MO-1gQFA.
\begin{Df}
An MO-1gQFA is a five-tuple ${\cal A}=( Q,\Sigma, q_1, \{{\cal E}_\sigma\}_{\sigma\in\Sigma},
Q_{a})$, where $Q$ is a finite  state set,  $\Sigma$ is a finite  alphabet, $q_1\in Q$ is the initial
state, ${\cal
E}_\sigma$ for each $\sigma\in\Sigma$ is a
 quantum operation having operation elements $\{E_{1}^\sigma,\cdots,E_{k}^\sigma\}$\footnote{It can be assumed that for all $\sigma$, the numbers of operation elements  of ${\cal
E}_\sigma$ are the same, since  we can get a maximal $k$ for all $\sigma$ and then add zero operator to those whose number of elements is less than $k$. } satisfying $\sum_{i=1}^k {E_i^\sigma}^\dagger E_i^\sigma=I$, and $Q_{a}\subseteq Q$ denotes the accepting state set, associated with a projector $P_{a}=\sum_{q\in Q_{a}}|q\rangle\langle q|$.\label{Df-MOQFA}
\end{Df}
 On scanning a symbol $\sigma$, the quantum operation ${\cal E}_\sigma$ is performed on the current state.
After  the operation corresponding to the last symbol is performed, a projective measurement is applied to determine acceptance.  Thus,  for the input string $x=\sigma_1\sigma_2\cdots \sigma_n\in\Sigma^*$,   ${\cal A}$ produces the  accepting probability  given by
\begin{align*}
P_{\cal A}(x)=\text{Tr}(P_{a}{\cal
E}_{\sigma_n}\circ\dots\circ{\cal E}_{\sigma_2}\circ{\cal
E}_{\sigma_1}(|q_1\rangle\langle q_1|)),
\end{align*}
where ${\cal E}_2\circ{\cal E}_1(\rho)$ stands for  ${\cal
E}_2({\cal E}_1(\rho))$.

 MO-1gQFA  have been thought to be a nice definition of QFA, since they possess nice closure properties and have a competitive computational power with their classical counterparts. The next lemma shows that any DFA can be simulated by a MQ-1gQFA.

 \begin{Lm}For any DFA ${\cal A}=(S,\Sigma,s_1,\delta,S_a)$, there exists an  MO-1gQFA  ${\cal A}'=(S,\Sigma,s_1,\{{\cal E}_\sigma\}_{\sigma\in\Sigma}, S_a)$ such that $|t\rangle\langle t|={\cal E}_{\sigma_n}\circ\cdots\circ{\cal E}_{\sigma_1}(|s\rangle \langle s|)$ iff $t=\delta^*(s, \sigma_1\cdots \sigma_n)$ where  $s, t\in S$ and $\sigma_1\cdots \sigma_n\in \Sigma^*$.
 \label{Lm:1}
\end{Lm}

{\it Proof.}   Let $E_s^\sigma=|\delta(s,\sigma)\rangle\langle s|$ where $s\in S$ and $\sigma\in\Sigma$. Then for each $\sigma\in\Sigma$, we get a quantum operation $\mathcal{E}_\sigma$ given by $\mathcal{E}_\sigma(\rho)=\sum_{s\in S} E_s^\sigma \rho {E_s^\sigma}^\dagger$. By a direct calculation, it is easy to verify that $|t\rangle\langle t|=\mathcal{E}_\sigma(|s\rangle\langle s|)$ iff $t=\delta(s,\sigma)$. Then the claimed result can be obtained by  induction on the length of the input string. \qed\\

From the above lemma, it can be seen that  $P_{{\cal A}'}(x)=1$ if ${\cal A}$ accepts $x$ and $P_{{\cal A}'}(x)=0$ if not. As a result, for each regular language there exists an MO-1gQFA which recognizes it  with certainty.
A full characterization of the languages recognized by MO-1gQFA is  as follows.
\begin{Lm} [\cite{LQ12}] (i) The languages recognized by MO-1gQFA are  regular languages; (ii) for each regular language, there exists an MO-1gQFA recognizing it with certainty.\label{lm:mo1gqfa}
\end{Lm}

\subsection{Simulation of CL-1QFA by MO-1gQFA}
In the following, we show that for a given  CL-1QFA, there exists an MO-1gQFA simulating it.
\begin{Th}Given a CL-1QFA  ${\cal
  A}$  with  control language ${\cal L}$ accepted by a DFA $\mathcal{A}'$, there exists an MO-1gQFA $\widehat{\cal A}$  simulating it exactly, with state set $Q\times S$ where $Q$ and $S$ are state sets of $\mathcal{A}$ and $\mathcal{A}'$, respectively.\label{th:cl1qfa}\end{Th}

{\it Proof.}   Let ${\cal
  A}=(Q, \Sigma, {\cal C},  q_1, \{U_\sigma\}_{\sigma\in\Sigma}, \mathcal{M}, {\cal L})$ be a  CL-1QFA.  As shown in Section \ref{sec:cl1qfa}, ${\cal A}$ is a hybrid model with a quantum component $Q$ and a classical component being a DFA $\mathcal{A}'=(S,{\cal C}, s_1,\delta,  S_a)$ accepting ${\cal L}$.  The  idea of the simulation is to encode the communication from $Q$ to $\mathcal{A}'$   into a controlled operation introduced in Section \ref{sec:co}.

 First by Lemma \ref{Lm:1}, there exists an MO-1gQFA   $\mathcal{A}''=(S,\mathcal{C},  s_1, \{\mathcal{F}_{c}\}_{c\in {\cal C}}, S_a)$,  in which each $\mathcal{F}_{c}$ on $\mathcal{H}_S$ is represented by operation elements $\{F^s_c\}_{s\in S}$
with $F^s_c=|\delta(s,c)\rangle \langle s|$, such that $|t\rangle\langle t|=\mathcal{F}_{c}(|s\rangle\langle s|)$ iff $t=\delta(s,c)$.

 Now we construct an MO-1gQFA $$\widehat{\cal A}=(\widehat{Q},\Sigma, \widehat{q}_1, \{\widehat{\mathcal{E}}_\sigma\}_{\sigma\in\Sigma}, \widehat{Q}_a)$$ from $\mathcal{A}$ and $\mathcal{A}''$ as follows:
\begin{itemize}
  \item  $\widehat{Q}=Q\times S$;
   \item $ \widehat{q}_1=(q_1,s_1)$;
\item $\widehat{Q}_a=Q\times S_a$, associated with the projector $\widehat{P}_a=I_Q\otimes\sum_{s\in S_a} |s\rangle\langle s|$ where $I_Q$ is the identity operator on ${\cal H}_Q$;
  \item for each $\sigma\in\Sigma$, $\widehat{\mathcal{E}}_{\sigma}$  has operation elements $\{ \widehat{E}_\sigma^{c,s}\}_{c\in \mathcal{C}, s\in S}$ where
      \[\widehat{E}_\sigma^{c,s}=P_cU_\sigma\otimes F_c^s=P_cU_\sigma\otimes |\delta(s,c)\rangle \langle s|.\]
 It is easy to verify that the collection $\{ \widehat{E}_\sigma^{c,s}\}$ satisfies the completeness condition.
Furthermore, for $\rho\otimes \varrho\in L(\mathcal{H}_Q\otimes \mathcal{H}_S)$, we have
\[\widehat{\mathcal{E}}_\sigma(\rho\otimes \varrho)=\sum_{c\in \mathcal{C}}P_cU_\sigma\rho U_\sigma^\dagger P_c\otimes \mathcal{F}_c(\varrho).\]
\end{itemize}

Now let us check the behavior of $\widehat{\cal A}$ on an input string. Suppose  $\widehat{\cal A}$ starts with the initial state $(q_1,s_1)$ and scans a symbol $\sigma$. Then the result state is
\begin{align*}\rho&=\widehat{\mathcal{E}}_\sigma(|q_1\rangle \langle q_1|\otimes |s_1\rangle \langle s_1|)\\
&=\sum_{c\in{\cal C}}P_{c}U_\sigma|q_1\rangle \langle q_1|U_\sigma^\dagger P_{c}\otimes\mathcal{ F}_{c}(|s_1\rangle\langle s_1|)\\
&=\sum_{c\in{\cal C}}P_{c}U_\sigma|q_1\rangle \langle q_1|U_\sigma^\dagger P_{c}\otimes |t_c\rangle\langle t_c|\end{align*}
where $t_c=\delta(s_1,c)$. In this way, after scanning a string $x=x_1x_2\cdots x_n\in\Sigma^*$,  the final state is
\begin{align*}\rho_x
&=\sum_{y\in{\cal C}^n}|\phi_y\rangle \langle\phi_y|\otimes |s_y\rangle\langle s_y|
\end{align*}
where  $|\phi_y\rangle=\prod_{i=1}^n(P_{y_i}U_{x_i})|q_1\rangle$ and $s_y=\delta^*(s_1,y)$.
 Note that $s_y\in S_a$ iff $y\in {\cal L}$.  Thus the  probability of $\widehat{\cal A}$ accepting $x$ is
\begin{align*}P_{\widehat{\cal A}}(x)&=\text{Tr}(\widehat{P}_a\rho_x)=\sum_{y\in{\cal C}^n}\chi_a(s_{y})\||\phi_y\rangle \|^2 \\
&=\sum_{y_1y_2\cdots y_n\in{\cal L}}\left\|\prod^{n}_{i=1}(P_{y_i}U_{x_i})|q_1\rangle\right\|^2,
 \end{align*}
where $\chi_a(s)$, defined after Eq. (\ref{prob:1qfac}), indicates whether $s$ is in $S_a$. Note that the above probability is equal to the one of ${\cal A}$ given in Eq. (\ref{f_CL}).
Therefore, we have completed the proof.\qed

\subsection{Simulation of 1QFAC by MO-1gQFA}
By a similar idea as before,  we can simulate 1QFAC by MO-1gQFA.
\begin{Th}
Given a 1QFAC with $Q$ and $S$ as its quantum and classical state sets, respectively, there exists an MO-1gQFA simulating it exactly, with $S\times Q$ as its state set.\label{th:1qfac}
\end{Th}

{\it Proof.} Let  ${\cal A}=(Q, S, \Sigma, q_1, s_{1}, \{ U_{s,\sigma}\}_{s\in S, \sigma\in \Sigma} , \delta,
 \{ {\cal M}_s\}_{s\in S})$ be a 1QFAC. As shown in Section \ref{sec:1qfac},  ${\cal A}$ is a hybrid model with a quantum component  $Q$ and a classical component being a DFA  ${\cal A}'=(S,\Sigma, s_1, \delta)$ without an accepting set. The  idea of the simulation is to encode the communication from  ${\cal A}'$ to $Q$  into a controlled operation introduced in Section \ref{sec:co}.

 First by Lemma \ref{Lm:1}, there exists an MO-1gQFA ${\cal A}''=(S,\Sigma, s_1, \{\mathcal{F}_\sigma\}_{\sigma\in\Sigma})$ (without an accepting set), such that $|t\rangle\langle t|=\mathcal{F}_\sigma(|s\rangle\langle s|)$ iff $t=\delta(s,\sigma)$.
Then we construct an MO-1gQFA  $$\widehat{\mathcal{A}}=(\widehat{Q},\Sigma, \widehat{q}_1, \{\widehat{\mathcal{E}}_\sigma\}_{\sigma\in\Sigma},\widehat{P}_a)$$ as follows:
 \begin{itemize}
   \item  $\widehat{Q}=S\times Q$;
   \item $\widehat{q}_1=(s_1,q_1)$;
   \item  For each $\sigma\in\Sigma$, $\widehat{\mathcal{E}}_\sigma$ on $\mathcal{H}_S\otimes \mathcal{H}_Q$ is given by
   $$\widehat{\mathcal{E}}_\sigma=(\mathcal{F}_\sigma\otimes \mathcal{I})\circ\mathcal{ E}_\sigma $$
   where $\mathcal{I}$ is the identity mapping on $L(\mathcal{H}_Q)$, and $\mathcal{ E}_\sigma$ is given by operation elements $\{E_s=|s\rangle\langle s|\otimes U_{s,\sigma}\}_{s\in S}$. Given $\rho\otimes\varrho\in L(\mathcal{H}_S\otimes \mathcal{H}_Q)$, we have
   $$\widehat{\mathcal{E}}_\sigma(\rho\otimes\varrho)=\sum_{s\in S}\langle s|\rho |s\rangle\mathcal{F}_\sigma(|s\rangle\langle s|) \otimes U_{s,\sigma}\varrho U_{s,\sigma}^\dagger,$$
which captures the idea that if the current classical state is $s$, then $U_{s,\sigma}$ is performed on the quantum component, and furthermore $s$ changes to another state according to $\mathcal{F}_\sigma$.
\item  Instead of specifying the set of accepting states,\footnote{Specifying an accepting state set is essentially equivalent to specifying a projective measurement.} we specify here the final projective measurement $\{\widehat{P}_a, \widehat{P}_r\}$.
Suppose   ${\cal M}_s=\{P_{s,a},P_{s,r}\}$. We let
\[\widehat{P}_a=\sum_{s_i\in S}|s_i\rangle\langle s_i| \otimes P_{s_i,a}\]
and
\[\widehat{P}_r=\sum_{s_i\in S}|s_i\rangle\langle s_i|\otimes P_{s_i,r}.\]
It is easily verified that $\widehat{P}_a$ and $\widehat{P}_r$ form a projective measurement.
 \end{itemize}

In the following we verify that $\widehat{\cal A}$  and  ${\cal A}$ have the same accepting probability for any given input string. Let $\rho_x$ be the state of  $\widehat{\cal A}$ after scanning $x$ and before  the final measurement. Then by induction on the length of the input string, it is easy to show that for each $x=x_1x_2\cdots x_n\in\Sigma^*$, we have
\begin{align*}
\rho_x=|s_{n+1}\rangle\langle s_{n+1}|\otimes|\psi_x\rangle\langle\psi_x|
\end{align*}
with
\begin{enumerate}
  \item [(1)] $|\psi_x\rangle=U_{s_n,x_n}\cdots U_{s_2,x_2}U_{s_1,x_1}|q_1\rangle$, and
  \item [(2)] $s_{i+1}=\delta(s_i,x_i)$ for $i=1,\cdots, n$.
 \end{enumerate}

Therefore,  the  probability of $\widehat{\mathcal{A}}$ accepting $x$ is
\begin{align*}
P_{\widehat{\cal A}}(x)&=\text{Tr}(\widehat{P}_a\rho_x)=\|P_{s_{n+1}, a}U_{s_n,x_n}\cdots U_{s_2,x_2}U_{s_1,x_1}|q_1\rangle\|^2,
\end{align*}
which is equal to the the accepting probability of ${\cal A}$  given in Eq. (\ref {prob:1qfac}). This completes the proof of Theorem \ref{th:1qfac}. \qed\\

\subsection{Simulation of 1QCFA by MO-1gQFA}
In this section we show how  1QCFA can be simulated by MO-1gQFA. This simulation is more complicated than previous simulations of CL-1QFA and 1QFAC, since now two-way communication occurs in  1QCFA.  However, the ideals are similar. \begin{Th}Given a 1QCFA ${\cal A}$ with $Q$ and $S$ as quantum and classical state sets, respectively, there exists an MO-1gQFA $\widehat{{\cal A}}$  simulating it exactly, with $Q\times S$ as the state set.\label{th:1qcfa}
\end{Th}

{\it Proof.} Let  $\mathcal{A}=(Q,S,\Sigma, {\cal C}, q_1, s_{1}, \{\Theta_{s,\sigma}\}_{s\in S,\sigma\in\Sigma},\delta,S_a)$ be a  1QCFA. As shown in Section \ref{sec:1QCFA},   $\mathcal{A}$ is a hybrid model comprising a quantum component  $Q$  and a classical component being a DFA $\mathcal{A}'=(S, \Sigma\times {\cal C}, s_1, \delta, S_a)$. The  idea of the simulation is to encode the communication between $Q$ and $A'$   into a controlled operation introduced in Section \ref{sec:co}.

First it follows from  Lemma \ref{Lm:1} that for the DFA $\mathcal{A}'$, there exists an MO-1gQFA
$\mathcal{A}''=(S,\Sigma\times {\cal C}, s_1,\{\mathcal{F}_{\sigma,c}\}_{\sigma,c\in \Sigma\times {\cal C}}, S_a),$
 such that $|t\rangle\langle t|=\mathcal{F}_{\sigma,c}(|s\rangle\langle s|)$ iff $t={\delta}(s, \sigma,c)$ for $(\sigma,c)\in \Sigma\times {\cal C}$.
 Then we construct an MO-1gQFA $$\widehat{\cal A}=(\widehat{Q},\Sigma,\widehat{q}_1, \{\widehat{\mathcal{E}}_{\sigma}\}_{\sigma\in\Sigma}, \widehat{Q}_a)$$  as follows:
\begin{itemize}
  \item  $\widehat{Q}=Q\times S$;
  \item $ \widehat{q}_1=(q_1,s_1)$;
 \item $\widehat{Q}_a=Q\times S_a$, associated with the projector $\widehat{P}_a=I_Q\otimes\sum_{s\in S_a}|s\rangle\langle s|$;
  \item Each $\widehat{\mathcal{E}}_{\sigma}$ on $\mathcal{H}_Q\otimes \mathcal{H}_S$ is given  by the following operation elements
\begin{align} \{ M^c_{s,\sigma}\otimes F^k_{\sigma,c}E_s: s\in S,  c\in \mathcal{C}, k\in K_{\sigma, c}\} \label{Eq-q} \end{align}
where: \begin{itemize}
 \item $E_s=|s\rangle \langle s|$ for $s\in S$, which is used to detect whether the classical state is $s$;
 \item  $\{M^c_{s,\sigma}\}_{ c\in \mathcal{C}}$ are measurement operators of $\Theta_{s,\sigma}$;
 \item  $\{F^k_{\sigma,c}\}_{k\in K_{\sigma,c}}$ are operation elements of $\mathcal{F}_{\sigma,c}$, that is, $\mathcal{F}_{\sigma,c}(\varrho)=\sum_{k\in K_{\sigma,c}}F^k_{\sigma,c}\varrho {F^k_{\sigma,c}}^\dag$.
      \end{itemize}
\end{itemize}

For each $\sigma\in \Sigma$, $\widehat{\mathcal{E}}_\sigma$ given by Eq. (\ref{Eq-q})  satisfies the completeness condition
  \begin{align*}
  &\sum_{s\in S}\sum_{c\in \mathcal{C}}\sum_{ k\in K_{\sigma,c}}(M^c_{s,\sigma}\otimes F^k_{\sigma,c}E_{s})^\dagger (M^c_{s,\sigma}\otimes F^k_{\sigma,c}E_{s})\\
  =&\sum_{s\in S}\sum_{c\in \mathcal{C}} {M^c_{s,\sigma}}^\dagger M^c_{s,\sigma}\otimes E_s^\dagger \left(\sum_{ k\in K_{\sigma,c}} {F^k_{\sigma,c}}^\dagger F_{c, k} \right) E_s\\
  =&\sum_{s\in S}\left(\sum_{c\in \mathcal{C}} {M^c_{s,\sigma}}^\dagger M_{s,c}\right)\otimes E_s^\dagger E_s\\
  =&I_Q\otimes\sum_{s\in S} E_s^\dagger E_s= I_Q\otimes I_S.
  \end{align*}
Furthermore, for $\rho\otimes\varrho \in L(\mathcal{H}_Q\otimes \mathcal{H}_S)$, by a direct calculation we have
\begin{align*}\widehat{\mathcal{E}}_{\sigma}(\rho\otimes\varrho)=\sum_{s\in S}\sum_{c\in \mathcal{C}}{M^c_{s,\sigma}}\rho {M^c_{s,\sigma}}^\dagger\otimes\langle s|\varrho |s\rangle \mathcal{F}_{\sigma,c}(|s\rangle \langle s|).\end{align*}
The above equation intuitively captures the idea that if the current classical state is $s$, then $\Theta_{s,\sigma}$ is performed on the quantum component; furthermore, if the outcome of $\Theta_{s,\sigma}$ is $c$, then $\mathcal{F}_{\sigma,c}$ is applied on the classical component.
For example, given the initial state $|q_1\rangle|s_1\rangle$, we have
\begin{align*}\widehat{\mathcal{E}}_{\sigma}(|q_1\rangle \langle q_1|\otimes|s_1\rangle\langle s_1|)=&\sum_{c\in \mathcal{C}}M^c_{s_1,\sigma}|q_1\rangle \langle q_1| {M^c_{s_1,\sigma}}^\dagger\otimes \mathcal{F}_{\sigma,c}(|s_1\rangle\langle s_1|)\\
=&\sum_{c\in \mathcal{C}}M^c_{s_1,\sigma}|q_1\rangle \langle q_1| {M^c_{s_1,\sigma}}^\dagger\otimes |{\delta}(s_1, \sigma,c)\rangle\langle {\delta}(s_1, \sigma,c)|.\end{align*}

In the following, we verify that the constructed MO-1gQFA $\widehat{\cal A}$ and the given 1QCFA ${\cal A}$ have the same accepting probability for any given input string. Let $\rho_x$ be the state of $\widehat{\cal A}$ after scanning $x$ and before the final measurement. Then for $x=x_1x_2\cdots x_n\in\Sigma^*$, we have
\begin{align}
\rho_x=\sum_{y\in {\cal C}^n}|\psi_y\rangle\langle \psi_y|\otimes|s_{n+1}\rangle\langle s_{n+1}| \label{state:simu1qcfa}
\end{align}
with
\begin{itemize}
  \item [(1)] $|\psi_y\rangle=M^{y_n}_{s_n, x_n}\cdots M^{y_2}_{s_2, x_2} M^{y_1}_{s_1, x_1}|q_1\rangle$, and
  \item [(2)] $s_{i+1}=\delta(s_i, x_i, y_i)$ for $i=1,\cdots, n$.
\end{itemize}

Actually, Eq. (\ref{state:simu1qcfa}) can be verified by induction on the length of  $x$ as follows.

{\it The basis step. } When $|x|=0$, that is, $x=\epsilon$, the result holds obviously.

{\it The induction step. } Suppose  the result holds for $x\in\Sigma^*$ with $|x|=n-1$. Then for $\sigma\in\Sigma$, we obtain
\begin{align*}
\rho_{x\sigma}&=\widehat{\mathcal{E}}_{\sigma}(\rho_x)=\sum_{y\in {\cal C}^{n-1}}\widehat{\mathcal{E}}_{\sigma}\left(|\psi_y\rangle\langle \psi_y|\otimes|s_{n}\rangle\langle s_{n}|\right)\\
&=\sum_{y\in {\cal C}^{n-1}}\sum_{y_n\in {\cal C}} M^{y_n}_{s_n, \sigma}|\psi_y\rangle\langle \psi_y| {M^{y_n}_{s_n, \sigma}}^\dagger \otimes{\cal F}_{\sigma,y_n}(|s_{n}\rangle\langle s_{n}|)\\
&=\sum_{yy_n\in {\cal C}^{n}}|\psi_{yy_n}\rangle\langle \psi_{yy_n}|\otimes |s_{n+1}\rangle\langle s_{n+1}|
\end{align*}
where $s_{n+1}=\delta(s_n,\sigma, y_n)$ and $|\psi_{yy_n}\rangle=M^{y_n}_{s_n, \sigma}|\psi_y\rangle$. Thus Eq. (\ref{state:simu1qcfa}) holds.

Now the probability of $\widehat{\cal A}$ accepting $x=x_1x_2\cdots x_n\in\Sigma^*$ is
\begin{align*}
P_{\widehat{\cal A}}(x)&=\text{Tr}(\widehat{P}_a\rho_x)=\sum_{y\in {\cal C}^n}\chi_a(s_{n+1})\| |\psi_y\rangle \|^2\\
&=\sum_{y\in {\cal C}^n}\chi_a(s_{n+1})\| M^{y_n}_{s_n, x_n}\cdots M^{y_2}_{s_2, x_2} M^{y_1}_{s_1, x_1}|q_1\rangle\|^2,
\end{align*}
which is equal to the probability of ${\cal A}$ accepting $x$ given in Eq. (\ref{prob:1qfac}).
\qed\\

\subsection{Language recognition power and equivalence problem of hybrid QFA}
In the study of QFA, an important problem is to characterize the language classes recognized by various models (e.g., \cite{AF98, BC01,BP99}). This problem was considered for CL-1QFA in \cite{Ber03, MP06} and for 1QFAC in \cite{QMS09}.
 Using our results presented in the previous sections, we can show that the three hybrid models indeed have the same language recognition power in the following sense.

\begin{figure}[tbp]\centering
\includegraphics[width=65mm]{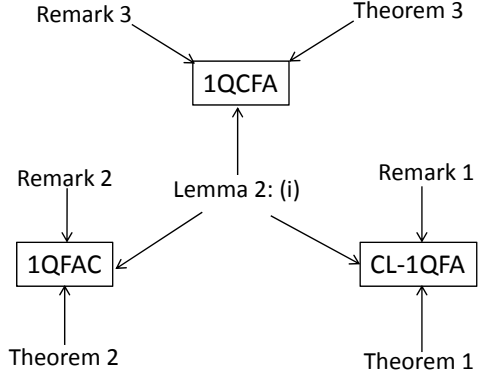}
\caption{A diagram illustrating the idea for proving  Corollary \ref{co:lang-power}. For example, the language recognition power of CL-1QFA follows from  item (i) of Lemma \ref{lm:mo1gqfa}, Remark \ref{rm:cl1qfa}, and Theorem \ref{th:cl1qfa}. }\label{Fig:lang-power}
\end{figure}

\begin{Co} The models
CL-1QFA, 1QFAC, and 1QCFA all recognize exactly the class of regular languages. \label{co:lang-power}
\end{Co}
{\it Proof.} The proof is depicted in Fig. \ref{Fig:lang-power} \qed\\

 Two QFA over the same input alphabet  are said to be equivalent if they have the same accepting probability for each input string.
The equivalence problem of a QFA model is that given any two automata of the  model, decide whether they are equivalent.  This problem has been proven to be decidable for  several quantum models \cite{LQ12, LQ06, LQ08a,  QZL09, QMS09}.  Similar problems were also discussed for
probabilistic automata \cite{Paz71,Tze92}. Specially,  it has been proven that two MO-1gQFA $\mathcal{A}_1$ and $\mathcal{A}_2$ are equivalent if and only if they have the same accepting probability for the input string with length no more than $n_1^2+n_2^2-1$ where $n_1$ and $n_2$ are  numbers of   states of $\mathcal{A}_1$ and $\mathcal{A}_2$, respectively (see \cite{LQ12}, Theorem 9).\footnote{In the original result, the bound was given by $(n_1+n_2)^2$, but it can be slightly improved to $n_1^2+n_2^2-1$ by a more careful analysis on the original proof.} This result, together with Theorems \ref{th:cl1qfa}, \ref{th:1qfac} and \ref{th:1qcfa}, immediately leads to the decidability of equivalence problem for hybrid QFA.
\begin{Co}Two 1QCFA (CL-1QFA, or 1QFAC) $\mathcal{A}_1$ and $\mathcal{A}_2$ are equivalent if and only if they have the same accepting probability for the input string with length no more than $(k_1n_1)^2+(k_2n_2)^2-1$ where $k_i$ and $n_i$ are  numbers of classical and quantum states of $\mathcal{A}_i$, respectively, $i=1,2$.
\end{Co}

\subsection{Equivalence between MO-1gQFA and ancilla QFA}
 In the previous section, it has been proved that  MO-1QFA can simulate several existing hybrid  QFA, which shows the generality of  MO-1gQFA.  It is, however, worth mentioning that  before Hirvensalo \cite{Hir08a, Hir08b} suggested this model, a model called ancilla QFA had already been proposed  by Paschen \cite{Pas00}. Although the two models were proposed with different motivations,  they actually represent the same model.
\begin{Df}
An ancilla QFA is a six-tuple ${\cal A} = (Q,\Sigma, \Omega, q_1, \delta,
 Q_a)$, where $Q$, $\Sigma$, $\Omega$  $q_1\in Q$ and  $Q_a\subseteq Q$ respectively  denote a finite state set, a finite input alphabet,  a finite output alphabet£¬ the initial state, and the accepting state set,
and the transition function $\delta$ : $Q\times \Sigma\times Q
\times\Omega ~\longrightarrow \mathbb{ C}$ satisfies
\begin{align}
\sum_{p\in Q,
\omega\in\Omega}\delta(q_1,\sigma,p,\omega)^*\delta(q_2,\sigma,p,\omega)=\begin{cases}1,&
q_1=q_2\\
0,&q_1\neq q_2 \end{cases}
\label{Acq}\end{align}
for all states $q_1, q_2\in Q$ and $\sigma\in\Sigma$.  Here $\mathbb{ C}$ is the set of complex numbers, and the $*$ operation in Eq. (\ref{Acq}) denotes the complex conjugate.\label{Def-AQFA}
\end{Df}

An ancilla QFA has a one-way tape head that reads symbols in the input tape from left to right and has an output tape where an output symbol is written at each step.  Given an input string $\sigma_1\cdots \sigma_n\in\Sigma^*$, it starts with the initial state $q_1$ and reads the first symbol $\sigma_1$. Then with amplitude $\delta(q_1,\sigma_1, p,\omega)$, it changes to state $p$ and write $\omega$ on its output tape. After that, the automaton moves its tape head right to read the next symbol. The above procedure continues until the last symbol $\sigma_n$ has been scanned. Finally, it checks whether the final state is in the set $Q_a$. If yes, then the input is  accepted; otherwise, it is rejected.

\begin{Rm}  Since the output tape is never read in the evolution  of ancilla QFA, it is  simply assumed that the output symbol is reset at every step, which thus results in the constant size of the output tape.
\end{Rm}

For $\sigma\in\Sigma$ and $\omega\in\Omega$, we define
\[V_{\sigma,\omega}=\sum_{q_i,q_j\in Q}\delta(q_j,\sigma,q_i,\omega)|q_i\rangle\langle q_j|\]
and
\[V_{\sigma}=\sum_{\omega\in\Omega}V_{\sigma,\omega}\otimes|\omega\rangle.\]
In matrix notations, $V_{\sigma,\omega}$ is a $|Q|\times |Q|$ matrix.
 Then, Eq. (\ref{Acq}) is equivalent to
\begin{align*}V^\dagger_{\sigma}V_{\sigma}=\sum_{\omega\in\Omega}V^\dagger_{\sigma,\omega}V_{\sigma,\omega}=I,\end{align*}
which means that $V_{\sigma}$ is an isometric operator from $\mathcal{H}_Q$ to  $\mathcal{H}_Q\otimes \mathcal{H}_\Omega$.

 Suppose now that the  machine is  in the  current state $\rho$. Then its state $\rho'$ after  scanning
$\sigma$ can be obtained by tracing over  $\Omega$ as follows:
\begin{align*}
\rho'=\text{Tr}_\Omega(V_\sigma\rho V_\sigma^\dagger)=\text{Tr}_\Omega\left(\sum_{\omega,\omega'\in\Omega}V_{\sigma,\omega}\rho V_{\sigma,\omega'}^\dagger\otimes|\omega\rangle\langle\omega'|\right)=\sum_{\omega\in\Omega}V_{\sigma,\omega}\rho V_{\sigma,\omega}^\dagger.
\end{align*}

Therefore, for each $\sigma\in \Sigma$, the evolution of the machine is characterized by a quantum operation $\mathcal{E}_\sigma$ that has operation elements $\{V_{\sigma,\omega}:\omega\in\Omega\}$. As a result, an ancilla QFA  is  an MO-1gQFA.

On the other hand, we can construct an ancilla QFA equivalent to a given  MO-1gQFA. Let ${\cal A}=( Q,\Sigma, q_1,\{{\cal E}_\sigma\}_{\sigma\in\Sigma},
Q_{a})$ be an MO-1gQFA, where ${\cal
E}_\sigma(\rho)=\Sigma_{i=1}^k E_{\sigma,i}\rho E_{\sigma,i}^\dagger$.
Now we  construct an ancilla QFA  ${\cal A}' = (Q,\Sigma, \Omega, q_1, \delta,
 Q_a)$ where $Q,\Sigma,q_1, Q_a$ are identical to the ones in ${\cal A}$, $\Omega=\{1,2,\cdots,k\}$, and $\delta$ is defined by $\delta(q_j,\sigma,q_i,\omega)=\langle q_i|E_{\sigma,\omega}|q_j\rangle$ for $q_i, q_j\in Q$, $\sigma\in\Sigma$ and $\omega\in \Omega$.
It is straightforward to verify that the states of ${\cal A}$ and ${\cal A}'$ are identical at every step  for any given input string.

The above argument leads to the following theorem.
\begin{Th}
MO-1gQFA and ancilla QFA can simulate each other.
\end{Th}

Interestingly, the idea behind ancilla QFA was also used in \cite{NY09} to construct quantum interactive proof systems (QIP systems, for short), although the authors did not define the notion of ancilla QFA explicitly.
They proved each regular language can be recognized by a QIP system that takes a 1QFA as a verifier (\cite{NY09}, Proposition 4.2). The QIP system is essentially an ancilla QFA in that  at each step the verifier  simulates a DFA's behavior and  writes its current state as  output and then the prover erases the ouput.

To conclude this section, we would like to point out the relationship between ancilla QFA and the model quantum sequential machines (QSM) studied in  \cite{Qiu02, LQ06}.
\begin{Df} A QSM is a five-tuple
${\cal A}=(S,\Sigma,\Omega,s_1,\delta)$, where $S$ is a finite set of
internal states,  $\Sigma$ and $\Omega$ are
finite input and output alphabets, respectively,   $s_1\in S$ is the initial state, and $\delta:
\Sigma \times S\times \Omega\times S\rightarrow \mathbb{C}$ is a transition
 function satisfying
\begin{align} \sum_{\omega\in \Omega, t\in
S}\delta(\sigma,s,\omega,t)\delta(\sigma,s',\omega,t)^*=\begin{cases}1&s=s',\\
0& s\neq s'\end{cases}
\end{align}
for all states $s,s'\in S$ and every $\sigma\in \Sigma$.\end{Df}
 Intuitively we interpret $\delta(\sigma,s,\omega,t)$ as the transition
amplitude that ${\cal A}$ prints $\omega$ and enters state $t$ after
scanning $\sigma$ in the current state $s$. Thus, given an input string $x=x_1\cdots x_n\in \Sigma^*$, QSM ${\cal A}$ prints $y=y_1\cdots y_n\in \Omega^*$ with a certain probability denoted by $p(y|x)$. For the model of QSM, attentions are usually paid to the probability  $p(y|x)$ instead of acceptance or rejection.

\begin{Rm}
Now, if some accepting states are assigned to QSM ${\cal A}$, and we no longer care  the output, but focus on the accepting probability of the input, then we get an ancilla QFA.
In a word, an ancilla QFA  is essentially a  quantum sequential machine assigned with some accepting states.
\end{Rm}

\section{Conclusions}\label{sec:con}

 In this paper we  investigate three   hybrid  models of QFA---CL-1QFA, 1QFAC,  and 1QCFA---which differ from other QFA models by consisting of two interactive components: a quantum one and a classical one. The contribution of this paper is twofold. (i) First, we  characterize   structures of these models in a uniform  framework:  each hybrid model can be seen as a two-component communication system with certain communication pattern. (ii) Second, we   clarify the relationship between the hybrid models and other   models. Specifically, we show that CL-1QFA, 1QFAC, and 1QCFA can all be simulated exactly by MO-1gQFA. Some results in the literature concerning the language recognition power and the equivalence problem of these hybrid models follow directly from these relationships. In addition, MO-1gQFA and another early proposed model called ancilla QFA  are shown to be equivalent.

\section*{Acknowledgements}
 Li thanks  Dr. A. Yakaryilmaz for his reply to a query in \cite{YC10}, and  Dr. M. Hirvensalo for he kindly sending us the electronic copy of reference \cite{Hir08b}. 

\begin{thebibliography}{ABCD}


\bibitem{AF98} {A. Ambainis and  R. Freivalds},  {\em One-way quantum finite automata: strengths, weaknesses and generalizations},
in Proceedings of the 39th Annual Symposium on Foundations of
Computer Science, IEEE Computer Society Press, 1998, pp. 332-341.

\bibitem{AI99} { M. Amano and K. Iwama, }  {\em Undecidability on Quantum Finite
Automata}, in  Proceedings of the 31st Annual ACM Symposium on
Theory of Computing, 1999,  pp. 368-375.


\bibitem{ANT02} { A. Ambainis, A. Nayak, A.  Ta-Shma, and U. Vazirani}, {\em Dense quantum coding and quantum
automata}, J. ACM,  49 (2002),  pp. 496-511.

\bibitem{AW02} { A. Ambainis and J. Watrous, } {\em Two-way finite automata with
quantum and classical states}, Theoret. Comput. Sci.,  287 (2002),
pp. 299-311.
\bibitem{BC01} { A. Bertoni and M. Carpentieri, } {\em Regular Languages Accepted by Quantum Automata}, Inform. and Comput.,
165 (2001), pp. 174-182.

\bibitem{Ber03}{ A. Bertoni, C.  Mereghetti, and B.  Palano}, {\em Quantum
Computing: 1-Way Quantum Automata}, in Proceedings of the 9th
International Conference on Developments in Language Theory,
Lecture Notes in Comput. Sci. 2710,  Springer-Verlag, Berlin,
2003, pp. 1-20.




\bibitem{BP99} { A. Brodsky and N. Pippenger}, {\em Characterizations of 1-way quantum finite automata}, SIAM J.  Comput.,
31 (2002), pp. 1456-1478.

\bibitem{CHTW} R. Cleve, P. Hoyer, B. Toner, and J. Watrous, {\em Consequences and limits of nonlocal strategies}, in Proceedings of the 19th IEEE Conference on Computational
Complexity, Amherst MA, 2004,  pp.236-249.





\bibitem{Hir08a} M.  Hirvensalo, {\em Various Aspects of Finite Quantum Automata}, in 12th International Conference on Developments in Language Theory (DLT 2008), Lecture Notes in Computer Science,  vol. 5257: 21-33, 2008.


\bibitem{Hir08b} { M. Hirvensalo},  {\em Quantum Automata with Open Time Evolution}, International
Journal of Natural Computing Research,  1(2010), pp. 70-85.




\bibitem{KW97} { A. Kondacs and  J. Watrous}, {\em On the power of finite state automata},  in Proceedings of the 38th IEEE Annual
Symposium on Foundations of Computer Science, 1997, IEEE Computer
Society, pp. 66-75.



\bibitem{LQ12}	 L. Z. Li and D. W. Qiu et al, {\em Characterizations of one-way general quantum finite automata}, Theoret. Comput. Sci., 419 (2012), pp. 73-91.
\bibitem{LQ06}{ L. Z. Li and  D. W. Qiu},  {\em  Determination of equivalence between quantum sequential
machines},  Theoret. Comput. Sci.,  358 (2006), pp. 65-74.

\bibitem{LQ08a}{ L. Z. Li and  D. W. Qiu},  {\em Determining the equivalence for one-way quantum finite
automata}, Theoret. Comput. Sci.,  403 (2008),  pp. 42-51.

\bibitem{LQ08b}{ L. Z. Li and  D. W. Qiu}, {\em A note on quantum sequential
machines}, Theoret. Comput. Sci.,  410 (2009), pp. 2529-2535.



\bibitem{MC00} {  C. Moore and J. P. Crutchfield}, {\em Quantum automata and quantum grammars},  Theoret. Comput. Sci.,
237 (2000), pp. 275-306.



\bibitem{MP06} {  C. Mereghetti and B. Palano}, {\em Quantum finite automata with control
language}, Theoretical Informatics and Applications,  40 (2006),
pp. 315-332.

\bibitem{NC00}  {  M. A. Nielsen  and I. L. Chuang},  {\em Quantum Computation
and Quantum Information}, Cambridge University Press, Cambridge,
2000.






\bibitem{NY09} H. Nishimura  and  T. Yamakami, {\em An application of quantum finite automata to interactive proof systems},
Journal of Computer and System Sciences,  75 (2009), pp. 255-269,


\bibitem{Pas00}  {  K. Paschen, } {\em Quantum finite automata using ancilla qubits},
Technical report, University of Karlsruhe, 2000.

\bibitem{Paz71} A. Paz, {\em Introduction to Probabilistic Automata}, Academic Press, New 	
York 1971. 	


\bibitem{QLMG12}  {  D. W. Qiu, L. Z. Li, P. Mateus  and J. Gruska},  {\em Quantum finite automata, chapter of Handbook on Finite State based Models and Applications},  editor(s): Jiacun Wang, CRC press, October 16, 2012



\bibitem{Qiu02}  {  D. W. Qiu},  {\em Characterization of Sequential Quantum Machines}, Internat. J. Theoret.,
Phys.  41 (2002),  pp. 811-822.




\bibitem{QZL09} {  D. W. Qiu,  L. Z. Li, X. F. Zou, P. Mateus and J. Gruska}, {\em Decidability of the Equivalence of Multi-Letter Quantum Finite
Automata}, Acta Informatica,  48 (2011), pp. 271-290.




\bibitem{QMS09}{  D. W. Qiu, L. Z. Li, P. Mateus and A. Sernadas}, {\em Exponentially more concise quantum recognition of non-RMM regular languages}, arXiv:0909.1428.

\bibitem{Tze92} W. G. Tzeng,  {\em A Polynomial-time Algorithm for the Equivalence of
Probabilistic Automata}, SIAM J. Comput., 21 (1992), pp. 216-227.

\bibitem{Wat03}  J. Watrous, {\em On the complexity of simulating space-bounded quantum computations}, Computational Complexity 12 (2003), pp. 48-84.

\bibitem{Sel04} P. Selinger.  {\em Towards a quantum programming language}, Mathematical Structures in Computer Science, 14(4): 527-586, 2004.


\bibitem{YC10} A. Yakaryilmaz and A.C. Cem Say,{\em  Unbounded-error quantum computation with small space bounds}, Inform. and Comput., 209 (2011), pp. 873-892.



\end{thebibliography}
\end{document}